%% file: main.tex
\documentclass[11pt]{article}
\usepackage[margin=1in]{geometry} 

\usepackage[T1]{fontenc} 
\usepackage{palatino} 
\usepackage[version=3]{mhchem}
\usepackage[bottom]{footmisc}
\usepackage{amsmath,authblk,hyperref,xr,graphicx}
\usepackage[super]{natbib} 
\usepackage[labelfont=bf]{caption}
\makeatletter
\newcommand*{\addFileDependency}[1]{
  \typeout{(#1)}
  \@addtofilelist{#1}
  \IfFileExists{#1}{}{\typeout{No file #1.}}
}
\makeatother

\begin{document}

\title{ReactCA: A Cellular Automaton for Predicting Phase Evolution in Solid-State Reactions}

\author[1,2]{Max C. Gallant}
\author[1]{Matthew J. McDermott}
\author[1,2]{Bryant Li}
\author[1,2]{Kristin A. Persson \thanks{Corresponding Author: kapersson@lbl.gov}
}

\affil[1]{Materials Sciences Division, Lawrence Berkeley National Laboratory, Berkeley, CA, 94720}
\affil[2]{Department of Materials Science and Engineering, University of California, Berkeley, CA, 94720}

\maketitle

\pagebreak

\begin{abstract}
New computational tools for solid-state synthesis recipe design are needed in order to accelerate the experimental realization of novel functional materials proposed by high-throughput materials discovery workflows. This work contributes a cellular automaton simulation framework (ReactCA) for predicting the time-dependent evolution of intermediate and product phases during solid-state reactions as a function of precursor choice and amount, reaction atmosphere, and heating profile. The simulation captures rudimentary kinetic effects, the effects of reactant particle spatial distribution, particle melting and reaction atmosphere. It achieves conservation of mass using a stochastic, asynchronous evolution rule and estimates reaction rates using density functional theory data from the Materials Project \cite{jain_commentary_2013} and machine learning estimators for the the melting point \cite{hong_melting_2022} and the vibrational entropy component of the Gibbs free energy \cite{bartel_physical_2018}. The resulting simulation framework allows for the prediction of the likely outcome of a reaction recipe before any experiments are performed. We analyze five experimental solid-state recipes for \ce{BaTiO3}, \ce{CaZrN2} and \ce{YMnO3} found in the literature to illustrate the performance of the model in capturing reaction pathways as a function of temperature, reaction selectivity and precursor choice. Our approach allows for straightforward comparison of predicted mass fractions of intermediates and products with experimental results.  This simulation framework presents a step toward \textit{in silico} synthesis recipe design and an easier way to optimize existing recipes, aid in the identification of intermediates and identify effective recipes for yet unrealized inorganic solids.
\end{abstract}

\pagebreak

\section{Introduction}

The solid-state, or ``ceramic'' method is a simple and ubiquitous technique for synthesizing inorganic crystalline solids in which powder precursors are heated to elevated reaction temperatures under controlled atmospheric conditions \cite{grover_solid_2021}. This method is used at both the small-scale (e.g., research laboratories attempting the synthesis of new materials) and large-scale (industrial manufacturing processes) to produce a wide variety of important functional materials such as battery cathodes (including \ce{LiMnPO4}\cite{nwachukwu_research_2023} and \ce{LiFePO4}\cite{kang_optimized_2008}), the ferroelectrics \ce{BaTiO3}\cite{buscaglia_solid-state_2008}, \ce{YMnO3}\cite{todd_yttrium_2019}, and \ce{BiFeO3}\cite{han_low-temperature_2017}, and many superconductors, including \ce{FeSe_{0.88}}\cite{onar_solid_2015}, \ce{YBa2Cu3O_{6+x}}\cite{miura_observing_2021}, and \ce{MgB2}\cite{shi_synthesis_2004}. Despite the ubiquity of the method, no conventional system for designing or modeling solid-state synthesis recipes exists. Instead, recipes have long been designed primarily using expert knowledge (e.g., precursor selection from a common library or via phase diagram analysis) and heuristic guidelines \cite{merkle_tammannrule_2005, garn_solid_1979}.

The difficulty in modeling solid-state synthesis reactions can be illustrated by drawing a comparison to organic molecular synthesis in which recipes can be generated by working backward from a desired product molecule to a set of known precursors via a series of mechanistically well-defined steps in a process known as retrosynthesis \cite{szymkuc_computer-assisted_2016}. In contrast, high-temperature solid-state synthesis proceeds by spontaneous thermodynamic reactions which lack clearly defined intermediates and reaction mechanisms.  However, enabled by the recent rise of high-throughput density functional theory (DFT) calculations \cite{jain_high-throughput_2011, doerr_htmd_2016} and the databases generated by them \cite{jain_commentary_2013, kirklin_open_2015, saal_materials_2013, curtarolo_aflow_2012}, several new automatable methods for designing solid-state synthesis recipes have emerged. These methods, which include measures for determining the synthesizability of a desired target \cite{aykol_thermodynamic_2018, sun_thermodynamic_2016}, metrics for comparing the selectivity of reaction recipes \cite{aykol_rational_2021, mcdermott_assessing_2023}, tools for extracting synthesis data directly from natural language \cite{he_similarity_2020, kononova_text-mined_2019, kim_materials_2017}, and reaction networks which identify thermodynamically favorable pathways between specified precursor and target materials \cite{mcdermott_graph-based_2021}, have yielded early success in guiding synthesis recipe design, despite being built on zero-temperature simulations of ordered crystalline structures. Furthermore, advances in autonomous synthesis have increased throughput for synthesis experiments \cite{szymanski_autonomous_2023, chen_navigating_2024, burger_mobile_2020} and motivated the development of synthesis design algorithms that utilize experimental results to improve their planning \cite{szymanski_autonomous_2023-1}. While each of these methods provides an element of recipe design guidance, none of them allows for the quantitative prediction of the time and temperature-resolved emergence and consumption of phases during the execution of a synthesis recipe.

Though no \textit{a priori} simulation exists for predicting the progression of solid-state synthesis reactions, other reaction classes have been captured by simulation methods that are not neatly transferable to the solid-state case. For example, the kinetic Monte Carlo method is frequently used to model the evolution of species in gas or liquid phase molecular reactions (often in conjunction with reaction rates calculated using transition state theory \cite{pechukas_transition_1981, harms_advances_2020}). This method assumes integer numbers of discrete particles that transform via reaction into other sets of discrete particles, and it assumes that these particles are available to interact with each other with no heed paid to their spatial arrangement \cite{gillespie_exact_1977}. These two assumptions do not hold for solid-state reactions; instead, solid phases transform in continuous amounts from reactant to product and reactant particles do not move as freely as they do in liquid phase reactions (barring the presence of a molten flux or gas transport). Surface reactions have been successfully modelled by lattice Monte Carlo simulations to determine heterogeneous catalytic behavior, but they explicitly treat the motion of individual atoms \cite{andersen_practical_2019}. This method is not feasible for modeling the evolution of the powder contents of a solid-state reaction vessel because the large number of atoms are involved (often on the order of $10^{10}$-$10^{20}$ atoms or more) leads to intractable computing requirements. Indeed, any atomistic method presents similar limitations. Finally, phase field models have been used to model ionic diffusion during solid-state metathesis reactions \cite{huang_phase-field_2023}, but these methods require significant assumptions about the form of the governing equations and explicitly known mobilities for each of the species involved. Such mobility values are not readily calculable nor are they available in existing materials databases.

In light of these challenges, we present in this work a simulation framework (ReactCA) that predicts the time-dependent, quantitative evolution of phases over the course of a prescribed solid-state reaction as a function of precursor ratio, heating profile, and reaction atmosphere. To achieve this, we leverage the cellular automaton (CA) formalism \cite{gardner_mathematical_1970} which offers a flexible framework for addressing the unique challenges posed by solid-state reactions. A CA is defined by a grid of sites, each of which is assigned a state value. At each step in the evolution of the automaton, the state in each site is updated according to its own current value and the states of the sites neighboring it (the ``neighborhood''). The specific nature of the state values and the rule governing evolution (the ``evolution rule'') can be chosen to best suit the simulation problem at hand. As a result of this flexibility, cellular automata have been used in materials science and chemistry to model a variety of processes, including grain growth, crystallization, and surface adsorption/desorption  \cite{menshutina_cellular_2020,sieradzki_perceptive_2013}. Due to the fundamentally spatial nature of the neighborhood and the flexibility of the evolution rule, the CA structure is a natural choice for modeling solid-state synthesis.

The simulation framework described here utilizes zero-temperature thermodynamic properties of ordered crystalline compounds from the Materials Project as its primary input data. Importantly, some of the compounds we simulate here (and many of those found in the Materials Project) can accommodate disorder, which would increase the entropy of these phases and affect the energetics of the reactions which form them, especially at higher temperatures. To date however, no existing  database of computed material properties contains rigorous representations of configurationally disordered materials and their entropy. As a result, for the ordered materials in the Materials Project, we consider only the vibrational entropy contribution to the Gibbs formation energies as estimated by the machine learning method of Bartel et al. \cite{bartel_physical_2018}. These estimates, in conjunction with the cellular automaton formalism, are used to capture the thermodynamic and spatial features of solid-state reactions in addition to some rudimentary kinetic effects based on machine learning estimates of phase melting points. Our framework offers new functionality in automated solid-state synthesis planning in that it enables the facile prediction of quantitative reaction outcomes \textit{a priori} as a function of temperature profile, reaction atmosphere and precursor choice. We envision our framework being used as a straightforward, easily implemented method 1) for testing hypothesized recipes before attempting them in the lab, 2) for implementing a digital twin in a autonomous laboratory designing its own synthesis recipes, and 3) for refining synthesis parameters when used in conjunction with optimization frameworks.

\section{Theory and Computation}
\subsection{Cellular Automaton Model}

\begin{figure}[ht!]
\centering
\includegraphics[width=\textwidth,keepaspectratio]{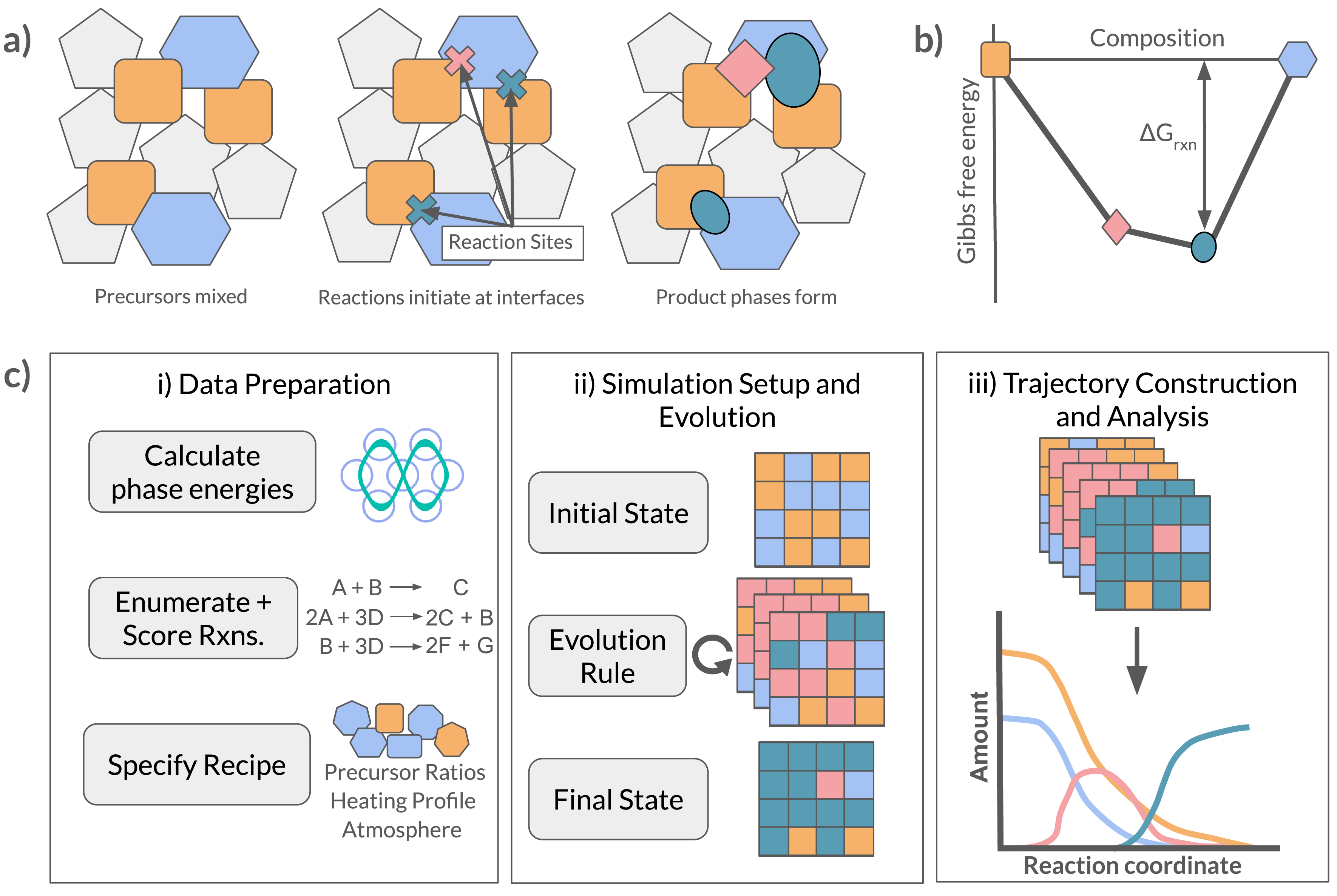}
\caption{\textbf{Modeling solid-state reactions with a cellular automaton}. a)  The progression of the initial stages of a solid-state reaction occurring via the pairwise interface reaction model; b) convex hull schematic representing the thermodynamics of reactions between two hypothetical solid precursor phases drawn as orange squares and light blue hexagons, with possible products given as the pink diamond and blue oval phases; c) schematic illustrating the main stages of the simulation: i) formation energies are obtained from the Materials Project and machine learning estimators are used to calculate the vibrational entropy part of the Gibbs energy of formation as a function of temperature and the melting point for each phase, reactions are enumerated and scored, and a recipe is specified which defines the desired precursors, a heating profile, and a reaction atmosphere; ii) a random initial arrangement of particles is generated, and the evolution rule is repeatedly applied to simulate the reaction; iii) simulation steps are concatenated into a trajectory which is analyzed to determine phase evolution over the reaction pathway.}
\label{fig:sim_overview}
\end{figure}

The solid-state reaction cellular automaton simulation described herein (ReactCA) is constructed based on the pairwise model of solid-state reactions. This model states that solid-state powder reactions proceed predominantly via sequential reactions at pairwise interfaces (i.e., between only two solid species at a time). This model has its theoretical basis in the spatial geometry of the contact regions between particles, and has been verified with \textit{in situ} experiments \cite{miura_observing_2021}. The solid-state reaction process, illustrated in Figure \ref{fig:sim_overview}a, proceeds via  the diffusion of atomic species driven by chemical potential differences across the interface between two reactant particles. As the reaction progresses, nuclei of one or more stable product phases form and grow at the interface, converting reactant material into product. Importantly, the local composition of the interface region is determined by the kinetic availability of reacting species and not constrained to reflect the overall composition of the precursor mixture. The first product phase to form is then a function of the ``local'' composition (as opposed to the overall composition) and the relative energetics of the possible product phases \cite{bianchini_interplay_2020}.

The thermodynamics of the pairwise interface reaction model are conveniently represented by the convex hull construction, in which the Gibbs free energy is shown as a function of the mixing ratio (i.e., mixture composition) of the two precursor phases (Figure \ref{fig:sim_overview}b). The interior points (i.e., pink diamond and blue oval) represent product phases that can form as the result of the reaction of the precursors. In two- or three-element systems, these points correspond to discrete compositions; however, for larger systems, the interior points can additionally correspond to balanced mixtures of two or more product phases. The vertical distance between the compositional axis and a product point is the change in free energy of the corresponding reaction. The geometry of the reaction hull contains information about the behavior of a particular reacting pair, as illustrated by reaction selectivity metrics based on thermodynamics developed in Ref. \citenum{mcdermott_assessing_2023}. This pairwise reaction model and the interface hull underpin the simulation described here, and in particular motivate the choice of the CA formalism, which naturally captures local interactions between neighboring entities.

The structure of the ReactCA simulation framework can be broken down into three stages (Figure \ref{fig:sim_overview}c). The first entails the automated collection and calculation of relevant phase thermodynamics, an assessment of a score function for estimating relative reaction rates, and the specification of the reaction recipe. In the second phase, an initial state (or arrangement of phases on a grid) is produced, then the evolution rule is repeatedly applied. Finally, the results of each application of the evolution rule are concatenated to form a trajectory which is analyzed to provide information about relative phase amounts at each time step.

\subsection{Phase Data Acquisition}

The input data for ReactCA is determined by the desired synthesis recipe which includes precursor ratios, a heating profile, and a reaction atmosphere (currently  gaseous atmospheres consisting of only a single element are supported, e.g. \ce{N2}, \ce{O2} or \ce{Ar}). The heating profile is defined by the user as a list of heating stages which each have a temperature and duration (specified by number of simulation steps). Once this recipe is defined, ordered crystal structures in the chemical system spanned by the reaction atmosphere and precursor phases are identified and their calculated formation energies are acquired from the Materials Project. Note that these formation energies are calculated via zero-temperature DFT while solid-state reactions occur at elevated temperatures. However, exact calculations of finite temperature formation energies are not available from existing high-throughput databases. To bypass this data deficiency, a machine learning descriptor given by Bartel et al.\cite{bartel_physical_2018} is used to estimate the vibrational entropy contribution to the Gibbs energy of formation for ordered solid phases at each of the temperatures specified in the reaction recipe. The Gibbs free energies of formation for common liquid/gaseous phases are acquired from experimental thermochemistry data (NIST-JANAF tables) \cite{chase_nist-janaf_1998}. Finally, the melting points of all phases are estimated using the graph neural network model from Hong et al.\cite{hong_melting_2022}.

\subsection{Reaction Enumeration and Scoring}

\begin{figure}[ht!]
\centering
\includegraphics[width=\columnwidth]{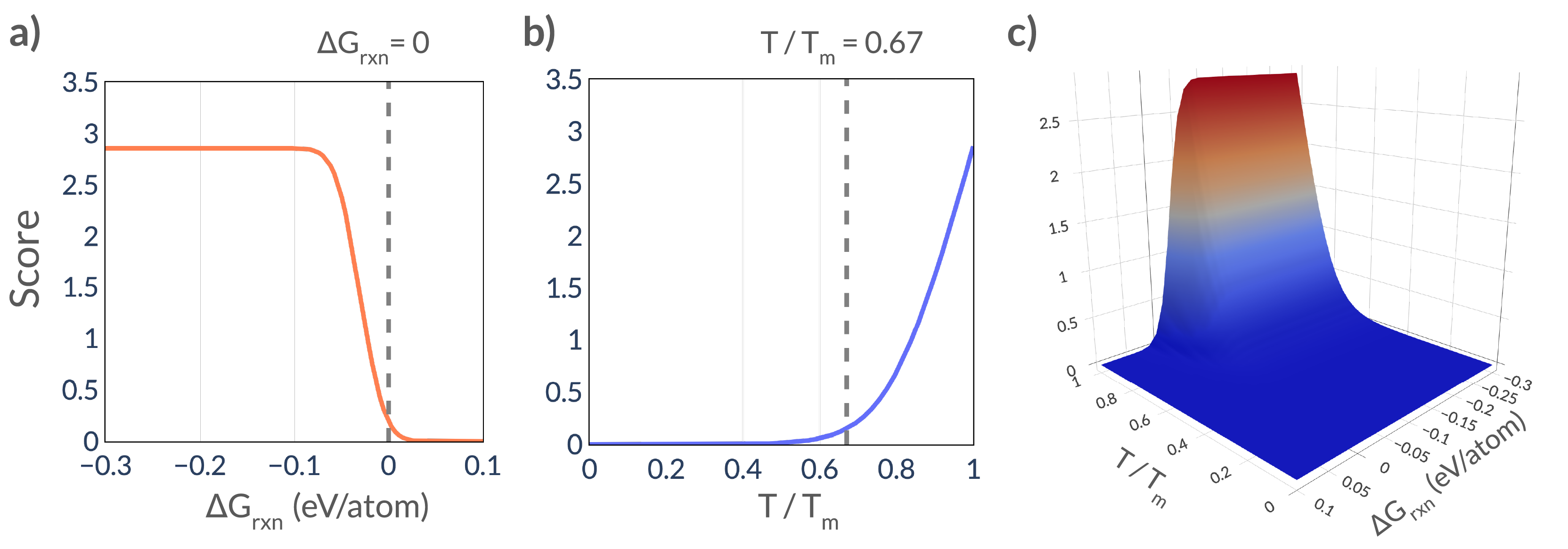}
\caption{\textbf{Scoring the likelihood of reactions as a function of reaction energy and temperature.} a) Score relationship with reaction energy; at a constant temperature above the Tamman temperature, endergonic reactions are vanishingly unlikely, while increasing exergonicity does yield an infinitely increasing reaction rate, b) Relation with temperature (assuming a constant, negative reaction energy); reaction likelihood increases quickly above the Tamman temperature, c) Score relationship plotted as a surface function of both inputs.}
\label{fig:score_fn}
\end{figure}

With the phases and energies acquired from the Materials Project, the \verb|reaction-network|\cite{mcdermott_graph-based_2021} Python package is used to identify all stoichiometrically possible reactions and calculate the changes in Gibbs free energy associated with them at each of the specified temperatures. While no general strategy exists for estimating the rate of solid-state reactions, predicting the evolution of phases during a reaction necessitates a model for relative reaction rates.  To accomplish this, a score, $S$, is calculated for each reaction at each temperature using a heuristic function (Eq. \ref{equation:score}), which returns the relative likelihood of each reaction occurring:

\begin{equation}
S = \frac{1}{2} ln\left[1 + \exp(a * (\frac{T_{\text{rxn}}}{T_{m,\text{reactant}}} - b))\right]*\left[1 + \text{erf}(-c (\Delta G_{\text{rxn}} + d))\right]
\label{equation:score}
\end{equation}

The score function described by Eq.~\ref{equation:score} is composed of two primary terms: a softplus function term and an error function term. The shape of this function (shown in Figure \ref{fig:score_fn}) captures: 1) the spontaneity of exergonic reactions, 2) the onset of reactions at temperatures equal to two-thirds of the melting point of the lowest melting point precursor (i.e., Tamman's Rule) and 3) the increase of reaction rate with temperature. The scaling parameters $a = 14$ and $b = 0.8$ were chosen to shift/scale the softplus function such that its ``onset'' is around the Tamman temperature ($\frac{T_{rxn}}{T_{m}} = \frac{2}/{3}$) and the parameters $c = 35$ and $d = 0.03$ were chosen to shift/scale the error function such that it is centered on the region just below $\Delta G = 0$ eV/atom. Other values near the ones shown here for $a, b, c$ and $d$ were experimented with, but the variations did not significantly alter the simulation outcomes.

The softplus function was chosen to encode Tamman's rule because of its ``soft'' activation (i.e., above the Tamman temperature). The error function was chosen to encode spontaneity because it behaves as a dial that abruptly ``ramps up'' for exergonic reactions. While these effects could also have been encoded using piecewise functions (e.g., a rectified linear unit in place of the softplus function, or Heaviside function in place of the error function), we opt for smooth alternatives which ``smear'' the onset of each effect over a range of values. This smearing allows for a degree of accommodation for uncertainty in our input Gibbs energy and melting point estimates. Most importantly, the scoring function can easily be updated to accommodate more sophisticated functionality, e.g. based on kinetics and local availability of reactive species. 

\subsection{Simulation Evolution}

After phase data is collected and reactions are enumerated and scored, an initial simulation state is defined. The simulation box for this automaton is a three-dimensional region of space subject to periodic boundary conditions and discretized into a grid of cubic cells. To establish an initial state, each cell is randomly assigned a phase occupancy according to the precursor ratios given by the reaction recipe along with a volume equal to 1.0. We assign no scale or unit to this value because only the \textit{ratio} of the volumes of neighboring cells is relevant to this simulation. This value is used in achieving conservation of mass and should not be interpreted as a literal measure of the physical extent of the simulation.

\begin{figure}[ht!]
\centering
\includegraphics[width=\textwidth,keepaspectratio]{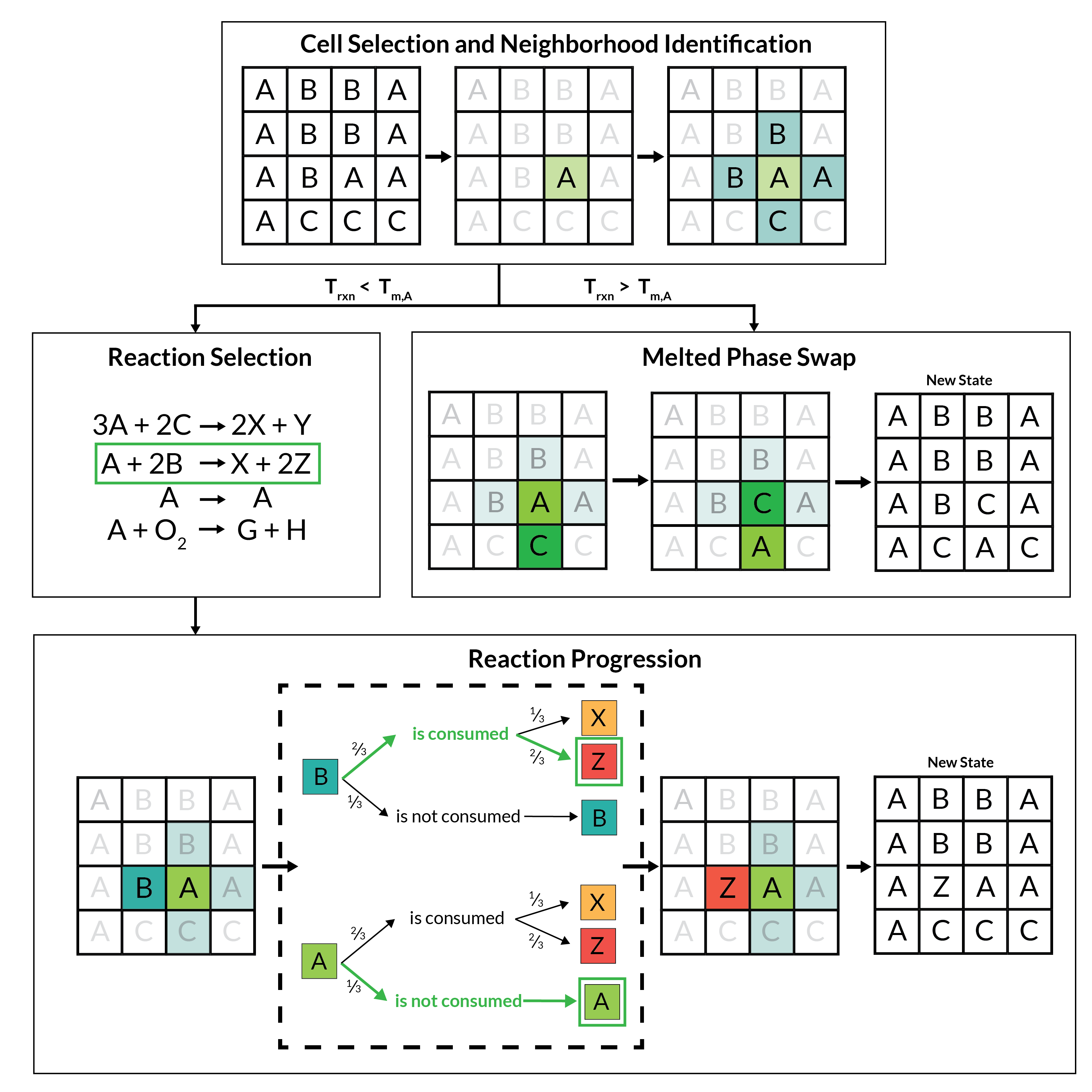}
\caption{\textbf{Evolution of phases in the simulation according to the evolution rule}. In the top panel a cell is randomly selected (pale green) from the simulation and its neighbors are identified (teal). Next, if the simulation temperature is above the melting point of the phase in the selected cell, the Melted Phase Swap action occurs (middle-right). If not, reactions between the selected cell and its neighbors are enumerated and a reaction is randomly chosen using the reaction scores as probabilities (middle-left). Finally, each of the reacting cells' phases are replaced (or not) according to probabilities given by the stoichiometric coefficients of the selected reaction. These probabilities are indicated by the small fractions decorating each arrow in the bottom panel. Note that the CA implemented in this work uses a three-dimensional simulation state, but only two dimensions are shown here for clarity.}
\label{fig:evolve_actions}
\end{figure}

The evolution rule determines the phase occupancy and volume value of the selected cell in the next simulation time step. It is applied to a single cell at a time, selected at random, meaning that this simulation is an asynchronous cellular automaton (as opposed to a standard cellular automaton, in which every cell is updated simultaneously) \cite{fates_guided_2013}. This rule ensures that one of two actions occurs: 1) a swap, or 2) a reaction. These actions are illustrated in Figure \ref{fig:evolve_actions}. The definitions for the simulation state and shape, along with the evolution rule were implemented using the \verb|pylattica| Python package\cite{gallant_pylattica_2024}.

\subsubsection{Action 1: Melted Phase Swap}

If the current reaction temperature is above the melting point of the phase in the selected cell, the state of the selected cell is swapped with one of its neighbors, chosen at random, as shown in Figure \ref{fig:evolve_actions}. To accommodate uncertainty in the estimation of the melting point, the onset of the likelihood of this swap is smeared over a range of relative temperature values. Specifically, the swap likelihood begins ramping up as a function of reaction temperature at $T_{rxn} = 0.8T_m$, increases to a 95\% probability when $T_{rxn} = T_m$, and reaches a 99\% probability at $T_{rxn} = 1.2T_m$. This behavior is shown graphically in Supplementary Figure \ref{supp_fig:swap_probability}. The swapping motion facilitates movement of the reaction vessel contents and can capture heightened reactant movement during flux-mediated reactions in which the presence of a liquid phase makes reactants more able to access each other. This is crucial for capturing more realistic reaction dynamics in many solid-state reactions.

\subsubsection{Action 2: Reaction Progression}

If the phase occupying the selected cell is determined to be solid (i.e., it has a melting point higher than the current temperature), a reaction is selected. In this step, the reaction library is consulted to identify possible reactions between the selected cell and each of its neighbors. Reactions between the phase in the selected cell and the reaction atmosphere are also considered. From this list of possible interactions, a reaction is chosen randomly with a probability that is proportional to its score obtained from Eq.~\ref{equation:score} As a result, reactions with higher scores occur more frequently than reactions with lower scores. This scheme has the net effect that higher-scored reactions proceed faster. Once a reaction is chosen, the reaction proceeds at each of the reacting cells; note that only a single cell is involved if the other reactant is contained in the atmosphere (e.g., gaseous \ce{O2}). This procedure entails several steps (Figure \ref{fig:evolve_actions}):
\begin{enumerate}
    \item A probability distribution over the reactants is constructed. The stoichiometric coefficients taken from the reaction are used as the weights in this distribution.
    \item A random draw from the resulting distribution is performed. If the resulting phase matches the phase of the reacting cell, the process proceeds to the next step. If it does not match, the step ends and the cell is left unchanged.
    \item If the reaction proceeds, a second distribution is constructed over the reaction products (again using their stoichiometric coefficients as weights).
    \item A draw from this distribution is used to select a product phase.
    \item The reacting phase is replaced with the product phase.
    \item The volume of the cell is scaled according to the ratio between the volume of the products and the reactants.
\end{enumerate}

Importantly, the process described above utilizes probability distributions over the stoichiometric coefficients of the reaction to maintain conservation of mass within the automaton. For a given reaction, the coefficients of the reactants provide the probability that each will be consumed during a given occurrence of that reaction. This ensures that the reactants are consumed at the correct rate relative to each other. The coefficients on the products of the reaction provide the probability that each one will be produced by a given occurrence of the reaction, similarly ensuring that the products of each reaction are produced at the correct rate relative to each other. Finally, scaling the volume of the simulation cell after its contents have been replaced during a reaction ensures that the correct amount of product phase is produced relative to the amount of reactant consumed by the reaction. A more detailed explanation of this process in conjunction with an example is provided in the Supplementary Information.

\subsection{Trajectory Construction and Analysis}

A simulation run typically entails hundreds of thousands of applications of the evolution rule described above. When the simulation is complete, the results are concatenated into a trajectory that can be analyzed to understand the reaction pathway as a series of steps and discrete intermediate species. Because ReactCA relies on random draws from probability distributions over the possible actions, reactions, and product phases, several simulations are run in parallel, each utilizing a different random starting state. Though the choice of starting state does not affect the qualitative outcome of the simulation, each run is characterized by differing fluctuations, and represents a unique sampling of the distributions in the automaton. As shown in Figures \ref{supp_fig:convergence_final_mass} and \ref{supp_fig:convergence_max_mass}, when a sufficiently large simulation box is used, the standard deviation of the maximum and final mass fractions attained by each phase across a set of parallel trajectories is reduced to less than 1\%. To construct the final result, the individual outputs of these parallel simulations are ensemble averaged to yield an overall trajectory.

\section{Results and Discussion}

To test the efficacy of ReactCA in describing real solid-state reactions, we apply it to model several case studies selected from the literature where high-quality \textit{in situ} phase evolution data is available. For each case, we compare the predicted and observed final products of the synthesis reaction as well as the appearance (or disappearance) of intermediate and impurity phases.

\subsection{Product Selectivity in \texorpdfstring{\ce{BaTiO3}}{TEXT} Recipes}

Barium titanate is a well-known multiferroic material with a significant body of synthesis literature. While there are a number of well-known recipes for producing this material, we refer to the recent solid-state reaction selectivity study of Ref. \citenum{mcdermott_assessing_2023}, which tested and compared nine different \ce{BaTiO3} synthesis recipes characterized over a range of temperatures with synchrotron X-ray diffraction (XRD). A selection of these recipes and the corresponding reactions are simulated here using ReactCA to illustrate the way reaction selectivity is expressed in a phase evolution prediction.

\begin{figure}[ht!]
\centering
\includegraphics[width=\textwidth,height=\textheight,keepaspectratio]{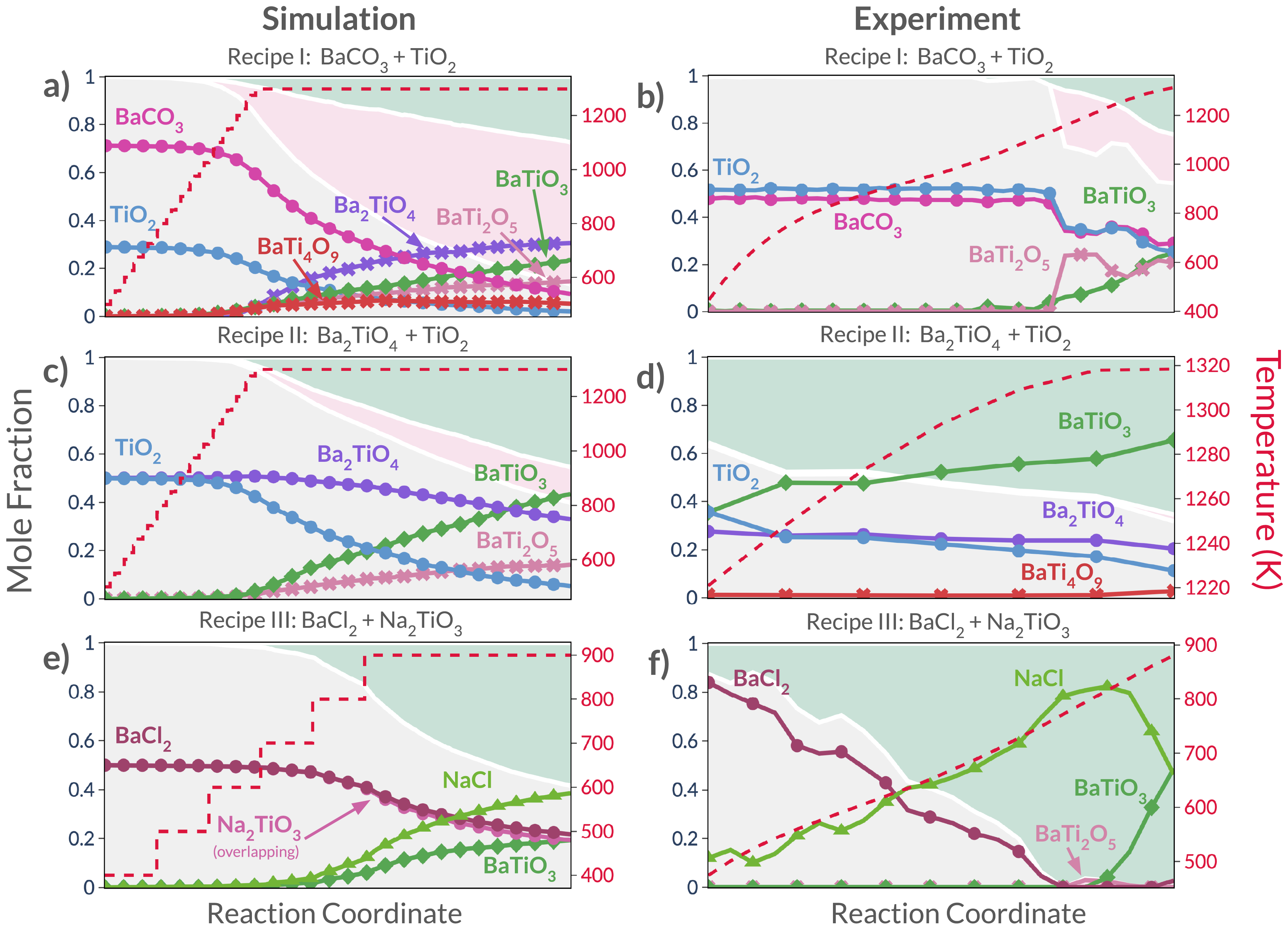}
\caption{\textbf{Simulated (left) and experimental (right) reaction evolution plots for the selected \ce{BaTiO3} recipes}. In each of these plots, the x-axis corresponds to the reaction coordinate, and the y-axis corresponds to mass fraction. The background of each plot is colored according to the amount of precursor (grey), impurity/intermediate (pink), and target/byproduct (light green) over the course of each simulation or experiment. Each of the traces represents the amount of each phase during the reaction. Traces marked with circles correspond to precursor phases, those marked with exes correspond to intermediate or impurity phases, and those marked with diamonds correspond to \ce{BaTiO3}, the expected target product phase. The dashed red lines show the heating profile used for each simulation or experiment.}
\label{fig:batio3_rxns}
\end{figure}

Selectivity was assessed in previous work according to two metrics: primary competition, which quantifies the likelihood of impurities forming from reaction of precursors, and secondary competition, which quantifies the likelihood of subsequent reactions consuming the desired products after they form. In the case of both of these metrics, a lower value corresponds to a more selective reaction, that is, one that is more likely to form only the desired product phase. To illustrate the way that selectivity presents itself in ReactCA simulations, three recipes were selected: the conventional recipe (Recipe I - \ce{BaCO3} and \ce{TiO2}), a recipe with improved selectivity but a lower overall driving force (Recipe II - \ce{Ba2TiO4} and \ce{TiO2}), and a metathesis reaction with excellent selectivity (Recipe III - \ce{BaCl2} and \ce{Na2TiO3}).

\textbf{Recipe I.} The simulation results for Recipe I are shown in Figure \ref{fig:batio3_rxns}a and the corresponding experimental outcome is shown in Figure \ref{fig:batio3_rxns}b. This reaction received a low primary competition score but a perfect (zero) secondary competition score, suggesting that there were competing phases that could form from the original precursors but that if the desired products were formed, they would be unlikely to be consumed by any secondary reactions. In the corresponding experiment from Ref. \citenum{mcdermott_assessing_2023}, the \ce{BaTi2O5} phase formed as an impurity in conjunction with the product, \ce{BaTiO3}, at around 1100~K. The result of the reaction automaton simulation for this recipe, shown in Figure \ref{fig:batio3_rxns}a, predicts the formation of the target phase \ce{BaTiO3} as well as the \ce{BaTi2O5} impurity phase at the same onset temperature (1100~K). However, it also predicts the appearance of two additional  phases (\ce{Ba2TiO4} and  \ce{BaTi4O9}). The three impurities grow at a similar rate to the desired product, \ce{BaTiO3}, an effect also seen in the experiment. This result illustrates the way that secondary competition appears in a reaction: competing phases form during the  reaction of the precursor materials, but once the desired product, \ce{BaTiO3}, is formed, it is not consumed by any subsequent reaction.

\textbf{Recipe II.} The second reaction chosen here was shown~\cite{mcdermott_assessing_2023} to improve the selectivity of the first at the cost of lowering the driving force of the reaction by choosing compositional members toward the interior of the complex hull as precursors. The ReactCA simulation (shown in Figure \ref{fig:batio3_rxns}c) predicts a majority \ce{BaTiO3} formation, and the formation of the impurity \ce{BaTi2O5}. In the corresponding experiment (shown in Figure \ref{fig:batio3_rxns}d), \ce{BaTi4O9} is observed as an impurity but the simulated \ce{BaTi2O5} is notably absent. This will be discussed below. 

\textbf{Recipe III.} The final \ce{BaTiO3} reaction selected for simulation is a metathesis reaction using \ce{NaTiO3} as the as the \ce{Ti} source and \ce{BaCl2} for the \ce{Ba} source. In the original work, this reaction was selected for its strong exergonicity, and its strong selectivity scores. The high selectivity of this reaction is on display in the ReactCA prediction, and the prediction (shown in Figure \ref{fig:batio3_rxns}e) is in good agreement with the experimental results (Figure \ref{fig:batio3_rxns}f) -- the dominant products are the intended metathesis products: \ce{BaTiO3} and \ce{NaCl}. The main discrepancy between the prediction and the experiment is that the automaton predicts no other Ba-Ti-O phase formation, while the experiment indicates appearance of \ce{BaTi2O5}, though it is only a trace amount.

Across these three reactions, the selectivity differences between the recipes are apparent in the results from the CA simulations. The size of the green regions in Figure \ref{fig:batio3_rxns} show the overall trend from low selectivity (in the case of Recipe I), to increased selectivity (in the case of Recipe II) and finally to perfect selectivity (in the case of Recipe III). We also note a tendency for ReactCA to predict the appearance of unobserved impurity phases (particularly in the case of Recipe I, shown in Figure \ref{fig:batio3_rxns}a). This is a result of the evolution rule which samples reactions based on their calculated rates. The effect is especially strong for the Ba-Ti-O chemical system which contains many phases with similar energetics and which have similar melting points (the two features which are used in ReactCA to calculate reaction rates). Finally, in the reactions shown here, we also highlight a tendency of the automaton to overpredict accumulation of Ba-rich Ba-Ti-O ternary phases. For example, in the Recipe I simulation the most prevalent byproduct is \ce{Ba2TiO4} (Ba-rich), but in the corresponding experiment it is \ce{BaTi2O5} (Ba-poor). In the Recipe II simulation \ce{BaTi2O5} (Ba-rich) is the primary impurity, but in the experiment only \ce{BaTi4O9} (Ba-poor) appears. Finally, in Recipe III, the simulation predicts the formation of pure \ce{BaTiO3} (Ba-rich) while in the experiment, small quantities of \ce{BaTi2O5} (Ba-poor) were also observed. In light of this observation, we hypothesize that - given the similar energetics of these compounds - the preferential formation of Ba-deficient compounds in experiments may be related to kinetics of the ionic species across the reaction interface ~\cite{beauger_synthesis_1983}. This discrepancy between simulation and experiment motivates future work to develop new reaction rate estimators using system-specific kinetic models, perhaps in conjunction with yet-unrealized high-throughput databases of kinetic calculations.

\subsection{Intermediate Identification in \texorpdfstring{\ce{CaZrN2}}{TEXT} Synthesis}

\begin{figure}[ht!]
\centering
\includegraphics[width=\textwidth,keepaspectratio]{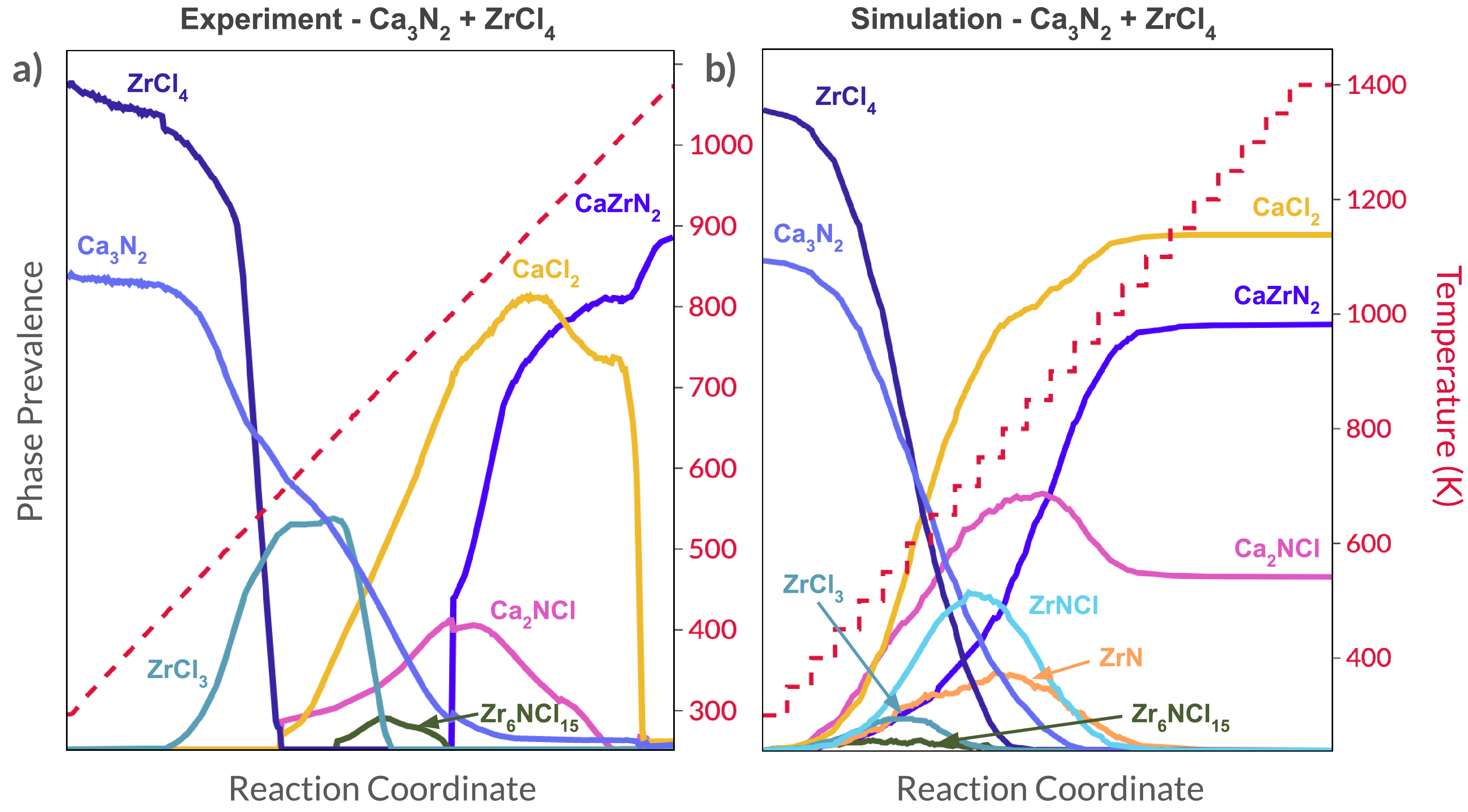}
\caption{\textbf{Resulting phase trajectories from simulation and experiment for \ce{CaZrN2} synthesis}. The experimental phase trajectories for the reaction using a) \ce{Ca3N2} and \ce{ZrCl4} as precursors and b) the simulated phase trajectories for the reaction using \ce{Ca3N2} and \ce{ZrCl4} as precursors. In a) the steep drop-offs of \ce{CaCl2} and \ce{ZrCl4} are caused by their melting and sublimating, respectively.}
\label{fig:cazrn_overall_rxn}
\end{figure}

Ternary nitride systems provide a wealth of materials discovery and synthesis opportunities. Recently, Rom et al. identified metathesis synthesis pathways that allowed them to produce the novel ternary nitrides \ce{CaZrN2} and \ce{CaHfN2} \cite{rom_mechanistically_2024}. Using \textit{in situ} XRD analysis, they constructed trajectories for each phase present during their synthesis reaction which used precursors \ce{Ca3N2} and \ce{ZrCl4} to yield \ce{CaZrN2}. This trajectory is reproduced in Figure \ref{fig:cazrn_overall_rxn}a. We performed a simulation for this reaction using a simulation box with side length of 15 cells, a heating profile consisting of a ramp phase from room temperature to 1400K, and an \ce{N2} reaction atmosphere. The resulting phase trajectories are shown alongside the experimental results from Ref. \citenum{rom_mechanistically_2024} in Figure \ref{fig:cazrn_overall_rxn}b. We note that in panel a) of this figure, the phase prevalence trace for \ce{Ca4Cl6O} is excluded. This phase appeared in the experiment due possibly to either impure precursor material or reaction with the quartz ampule \cite{rom_mechanistically_2024}, two effects which certainly represent practical synthesis considerations but are not relevant to the ideal environment represented by the automaton.

This simulation successfully captures the reaction pathway present in the experiment. The first intermediate phase to reach its peak in the simulation is \ce{ZrCl3}, which agrees with the early reduction of \ce{ZrCl4} to \ce{ZrCl3} in the experiment. The two intermediates which then follow in the experiment (a small amount of \ce{Zr6NCl15} and significant \ce{Ca2NCl}) are also present in the simulation with the same relative prevalences. Finally, the simulation predicts the major product (\ce{CaZrN2}) and byproduct (\ce{CaCl2}) with confidence, though some unreacted \ce{Ca2NCl} also remains. In addition to this pathway, the simulation predicts the appearance of two phases which are not observed in the experiments, \ce{ZrNCl} and \ce{ZrN}. In the following analysis we discuss the mechanisms by which these phases appear in the simulation and explain why those mechanisms may not have been active during the experiment. 

\ce{ZrNCl} appears in the simulation at roughly the same stage in the temperature trajectory as \ce{Ca2NCl} and the three reactions which facilitated the bulk of its formation are shown in Table \ref{tab:zrncl_rxns}. The most frequent of these reactions, Reaction (1), may have incorrectly occurred in the simulation because ReactCA, in its current form, does not include sublimation. Reaction (1) consumes \ce{Ca2NCl}, which only forms in the experiment (Figure \ref{fig:cazrn_overall_rxn}a) when the temperature has reached 600K and \ce{ZrCl4}, the other reactant, has sublimated. As a result, the two reactants may never have been sufficiently available to one another for this reaction to occur in the experiment. In contrast, since ReactCA has no method for estimating sublimation temperatures, \ce{ZrCl4} remains available when \ce{Ca2NCl} appears, allowing Reaction (1) to proceed. This hypothesis is supported by another experiment by Rom et al. in which \ce{Ca2NCl} and \ce{ZrCl4} were reacted directly as the initial solid precursors. In that experiment, both phases are present as solids and \ce{ZrNCl} appears as a prominent intermediate \cite{rom_mechanistically_2024}, suggesting that Reaction (1) does occur if both precursors are present.
\begin{center}
\begin{table}[ht]
    \centering
    \begin{tabular}{|c|c|c|c|}
     \hline
     Reaction & Score (600K) & $\Delta$ $\Theta_{rxn}$ $\frac{\text{eV}}{\text{atom}}$ & Occurrences \\ [0.5ex] 
     \hline
     (1) \ce{Ca2NCl + ZrCl4 -> 2CaCl2 + ZrNCl} & 0.165 & -0.562 & 6683 \\ 
     \hline
     (2) 2\ce{Ca3N2 + ZrCl4 -> ZrNCl + 3Ca2NCl} & 0.165 & -0.636 & 5103 \\
     \hline
     (3) \ce{Ca3N2 + 2ZrCl4 -> 3CaCl2 + 2ZrNCl} & 0.165 & -0.788 & 4533 \\
     \hline
    \end{tabular}
    \caption{The most frequently occurring \ce{ZrNCl}-forming reactions in the automaton simulation shown in Figure \ref{fig:cazrn_overall_rxn}b. We use the notation $\Delta \Theta_{rxn}$ as opposed to $\Delta G_{rxn}$ to indicate that the relevant thermodynamic potential in this system is a grand potential with \ce{N2} as the open species.}
    \label{tab:zrncl_rxns}
\end{table}
\end{center}
Reactions (2) and (3) in Table \ref{tab:zrncl_rxns} consume the same precursors: \ce{Ca3N2} and \ce{ZrCl4}. In the experiment however, Rom et al. propose that these phases instead react according to the following reaction:
\begin{align*}
    \ce{Ca3N2 + 6ZrCl4 &-> 6ZrCl3 + N2 + 3CaCl2} \\
    ~S = 0.165, ~\Delta \Theta_{rxn} &= -0.194~\tfrac{\text{eV}}{\text{atom}}~(T = 600K)
\end{align*}

This reaction and the three reactions in Table \ref{tab:zrncl_rxns} are all assigned the same score because they are highly exergonic (the thermodynamic component of the score function is maximized for all of them) and they share a lowest melting point precursor, \ce{ZrCl4} (so the melting point component of the score takes on the same value for all of them). Consequently, ReactCA does not differentiate the rates of these reactions, and they all occur with similar frequencies during the simulation. It is surprising that evidence for neither Reaction (2) nor Reaction (3) appears in the experiment, given their energetics (they are even more exergonic than the reaction proposed by Rom et al.) but there may be important differences in the kinetic accessibility of their product phases. In particular, the formation of the two binaries, \ce{ZrCl3} and \ce{CaCl2}, (along with the release of gaseous \ce{N2}) from these two compositionally dissimilar precursors may be more kinetically facile than the formation of the nitrogen-containing ternary phases \ce{ZrNCl} and \ce{Ca2NCl}. In support of this hypothesis, we note that Rom et al. observe the formation of a small amount of \ce{Zr6NCl15} (Figure \ref{fig:cazrn_overall_rxn}a), which could be interpreted as an incomplete incorporation of nitrogen while transforming of  \ce{ZrCl3} into \ce{ZrNCl}. Additionally while \ce{Ca2NCl} does appear in the experiment, Rom et al. suggest that its formation is facilitated by the reaction of more compositionally similar binaries, \ce{CaCl2} and \ce{Ca3N2} (which is precisely how it is formed in the simulation). When \ce{ZrNCl} does form, it is in the second experiment performed by Rom et al. (reacting \ce{Ca2NCl} with \ce{ZrCl4}). In this case, it may be that the presence of one ternary nitride (\ce{Ca2NCl}) leads to more facile formation of the other (\ce{ZrNCl}), potentially by providing more favorable nucleation sites on account of the similar layered structures and shared R$\overline{3}$m space group of the two phases. Considering these observations, the absence of \ce{ZrNCl} in the experiment strongly motivates the development of improved kinetic estimations for reaction rates in ReactCA and suggests that such estimations might be based in part on compositional or structural features of the precursors.

The simulation also predicts the emergence and consumption of \ce{ZrN}, an impurity that is not measured in either of the experiments. In their discussion, however, Rom et al. describe the growth of the product, \ce{CaZrN2}, as facilitated by the the slow growth of off-stoichiometric \ce{Ca_{x}Zr_{2-x}N2} starting from the \ce{ZrN} rocksalt phase. In other words, the early product phase in their experiments \textit{is} generated from \ce{ZrN}, but the material at that point is likely heavily defective and stoichiometrically ambiguous, which may be the reason that no explicit \ce{ZrN} phase appears in the XRD characterizations of the experiments. In contrast, no such defective or off-stoichiometric phase can be represented by ReactCA (which is limited to the ordered, crystalline, stoichiometrically-exact phases currently available in the Materials Project). As a result, the explicit appearance and disappearance of crystalline \ce{ZrN} is the best model the current version of the simulation can produce to represent the complex, continuous transformation in the experiment.

\subsection{Recovery of Observed Reaction Pathways in \texorpdfstring{\ce{YMnO3}}{TEXT} Synthesis}

The multiferroic \ce{YMnO3} has been the recent focus of a number of synthesis investigations into the effect of precursor selection, reaction atmosphere, and reaction temperature on both reaction pathways and also on the identity of the dominant product \cite{todd_yttrium_2019, mcdermott_graph-based_2021, todd_selectivity_2021}. In the first of these studies, Todd et al. propose reaction pathways at work during the formation of \ce{YMnO3} in a flux-assisted metathesis reaction and explain the lower reaction temperature required by their recipe in terms of the interplay between these pathways \cite{todd_yttrium_2019}. Building on this work, McDermott et al. were able to confirm using a reaction network that the suggested pathways were thermodynamically predicted by data within the Materials Project \cite{mcdermott_graph-based_2021}. We use this example here to illustrate the ability of ReactCA to predict temperature-dependent reaction pathways and intermediate and product mass fractions and to provide insight into the interactions between simultaneously occurring pathways.

\begin{figure}[ht!]
\centering
\includegraphics[height=400pt,keepaspectratio]{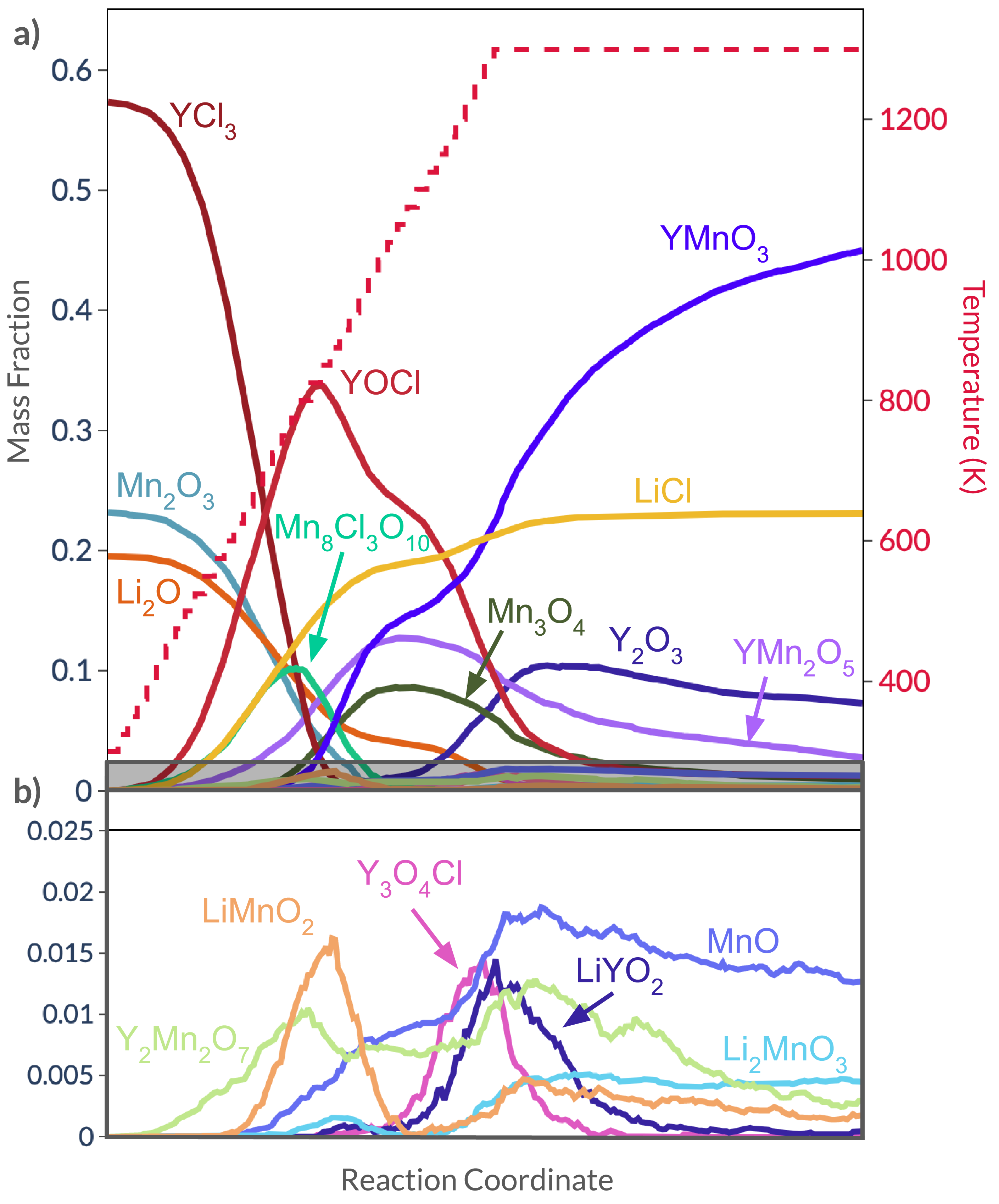}
\caption{\textbf{Resulting trajectory from simulation of the \ce{Li2CO3}-\ce{YCl3}-\ce{Mn2O3} reaction}. a) The total trajectory, b) a magnified view of all low-prevalence phases ($\leq$ 2.5\% by mass).}
\label{fig:ymno3_overall_rxn}
\end{figure}

The simulation for this reaction was configured to use a simulation box with a side length of 15 cells and a heating profile consisting of a ramp phase to 1300K followed by a hold phase at 1300K. We note that the peak temperature used here is slightly higher than the experimental maximum temperature of 1100K. This choice was made in order to accommodate uncertainty in the melting point and vibrational entropy estimates. A view of the resulting trajectory for this simulation is shown in Figure \ref{fig:ymno3_overall_rxn}a and a longer trajectory which includes the stabilization of the product phases is available in Supplementary Figure \ref{supp_fig:long_ymno3_traj}. The overall result shown here predicts \ce{YMnO3} as the dominant product phase and a number of intermediate phases, including \ce{YOCl}, \ce{Mn8Cl3O10}, \ce{YMn2O5}, and \ce{Mn3O4}. In Figure \ref{fig:ymno3_overall_rxn}b, a magnified view of the bottom of the trajectory is shown for for those phases which never formed greater than 2.5\% of the overall mass content in the simulation box. The multitude of phases present in this result illustrate the way that the reaction automaton samples many possible reaction pathways that can be traced from the initial precursor set.

\begin{figure}[ht!]
\centering
\includegraphics[width=\textwidth,keepaspectratio]{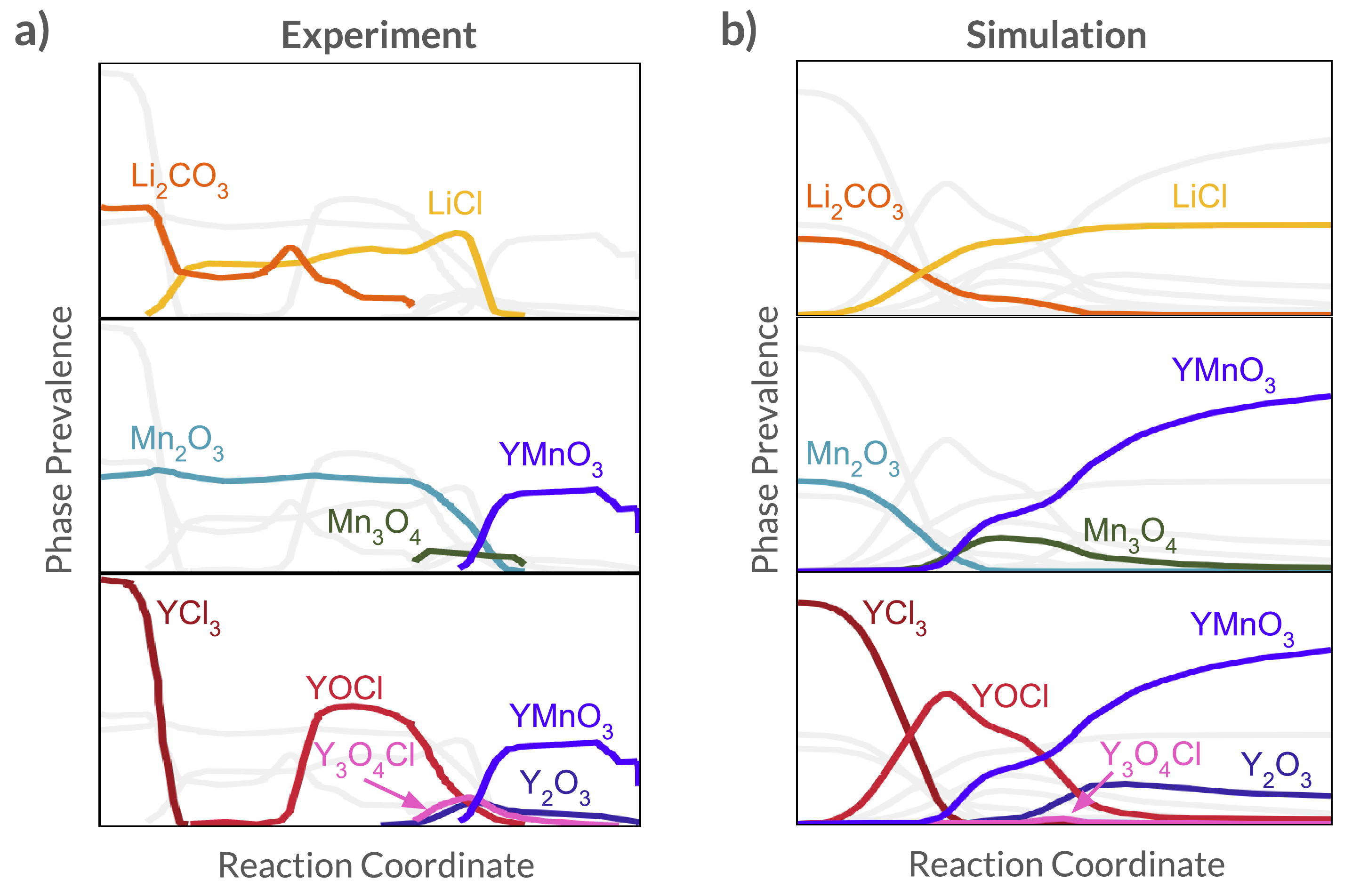}
\caption{Reaction pathways extracted from both a) the experimental synthesis results from Todd et al. \cite{todd_yttrium_2019} and b) the simulation trajectory of the reaction between \ce{Li2CO3}, \ce{YCl3} and \ce{Mn2O3}. The top panels illustrates the early emergence and subsequent plateau of \ce{LiCl}, the middle panels capture the emergence and then recession of the reduced \ce{Mn3O4} phase and the bottom panel shows the ordering and relative prevalence of the three key Y-O-Cl intermediates. Note that \ce{LiCl} disappears from the experimental phase trajectory in a) (which was generated using XRD data) only because it melts.}
\label{fig:ymno3_pathways}
\end{figure}

Amidst the complexity shown in Figure \ref{fig:ymno3_overall_rxn}, the reaction pathways identified by Todd et al. are remarkably well predicted by the ReactCA simulation. From XRD refinements, Todd et al. calculated the trajectory of each intermediate phase, and plotted them in groups according to cation. We reproduce these plots for the experimental data alongside similar plots generated from the simulation data in Figure \ref{fig:ymno3_pathways}. Every intermediate phase identified by Todd et al. is predicted by the reaction automaton. The relative amounts of these phases as well as the order of their appearance are also predicted with good accuracy, though \ce{Y3O4Cl} appears in only trace quantities in the simulation. However, we note that previous work highlights the high degree to which both \ce{Y3O4Cl} and \ce{YOCl} accommodate defects and disorder \cite{todd_defect-accommodating_2020}. Indeed, in another study Todd et al. assert that the transformation between these phases proceeds through an off-stoichiometric \ce{YO_{1+\epsilon}Cl_{1-\epsilon}} phase \cite{todd_defect-accommodating_2020}. Because our input data is limited to only the ordered, perfectly crystalline phases present in the Materials Project we do not include the effects of defects and disorder. As a result, we neglect the likely significant contribution of configurational entropy to the stability of these phases. This omission may explain the underestimation of \ce{Y3O4Cl} in this simulation result.

In addition to prediction of reaction intermediates, the ReactCA trajectory contains information about the specific reactions which yielded each phase. Of particular interest here is the formation mechanism for the product phase, \ce{YMnO3}. Todd et al. propose dual mechanisms for the formation of this phase - first, the faster, lower temperature ternary metathesis reaction between \ce{LiMnO2} and \ce{YOCl}, and second, the higher temperature reaction between \ce{Mn2O3} and \ce{Y2O3} (the latter of which is formed in part by consumption of \ce{YOCl}). Within the ReactCA trajectory, we identify two major classes of reactions that align with these two proposed reaction pathways. The first of these classes (Class I) is the reaction of \ce{LiMnO2} with one of a number of yttrium-containing intermediates (the most frequently occurring of which is \ce{YOCl}, followed by \ce{YCl3}). These reactions correspond to the low temperature, ternary metathesis step. The second of the two classes of reactions (Class II) producing \ce{YMnO3} are reactions between the refractory \ce{Y2O3} and either \ce{Mn2O3} and \ce{Mn3O4}. That reactions in the first class occur with any substantial frequency in this simulation is a surprising finding because the total \textit{amount} of \ce{LiMnO2} never exceeds 3\% by mass, suggesting that the phase is consumed at nearly a rate nearly equal to that at which it is produced. This may be a reasonable prediction, however, because the failure of this phase to accumulate in our simulation agrees strongly with the experiment performed by Todd et al, in which \ce{LiMnO2} appears in only trace quantities (its XRD pattern is poorly resolved from \ce{Mn3O4} which implies that the data in Figure \ref{fig:ymno3_pathways}a suggests the appearance of only small quantities of both \ce{Mn3O4} and \ce{LiMnO2} in the experiment).

\begin{figure}[ht!]
\centering
\includegraphics[width=\textwidth,keepaspectratio]{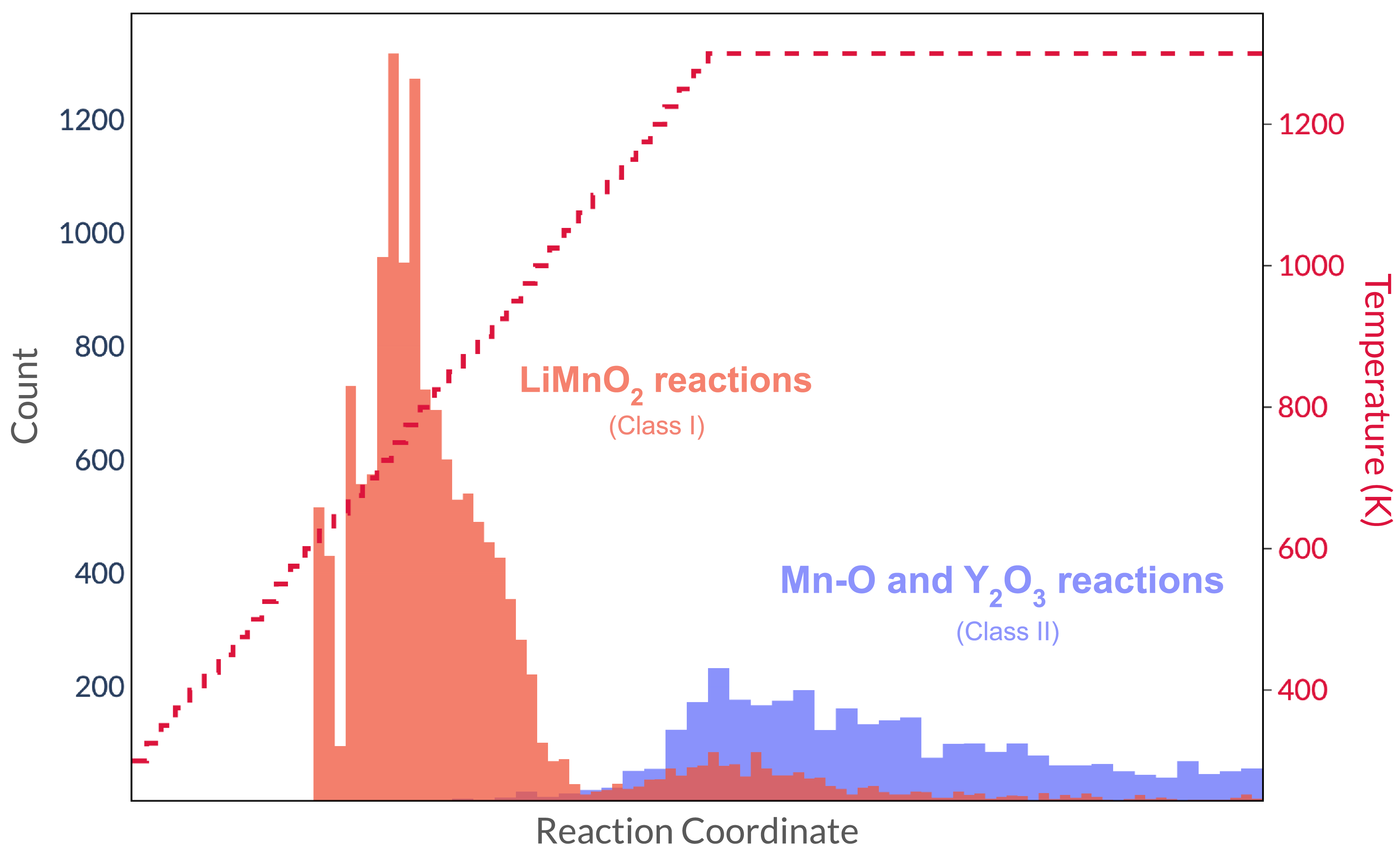}
\caption{Histogram illustrating the dominance of different classes of \ce{YMnO3}-forming reactions over the course of the simulation. The counts shown in red correspond to reactions which involve the \ce{LiMnO2} intermediate (the ternary metathesis route, i.e. Class I), and the counts in blue show reactions between the refractory \ce{Y2O3} and phases in the Mn-O chemical system (i.e. Class II).}
\label{fig:ymno3_rxn_hist}
\end{figure}

By counting the frequency of each of these classes of reactions as a function of the reaction coordinate, we illustrate in Figure \ref{fig:ymno3_rxn_hist} that the first ternary metathesis reaction class dominates early in the simulation at lower temperatures, and that the second reaction class which consumes the refractory \ce{Y2O3} occurs later in the simulation at higher temperatures. This result reflects the mechanism that Todd et al. proposed based on their experimental observations - ternary metathesis dominates at lower temperatures, and \ce{Y2O3} reactions proceed at higher temperatures, and indeed both mechanisms contribute to the formation of \ce{YMnO3}.

We emphasize that the analysis presented here represents an important increase in capability over the previous reaction network approach \cite{mcdermott_graph-based_2021}. In particular, the reaction network requires a set of expected product phases as input while ReactCA predicts the product without any prior target input. Additionally, the reaction network identified isolated pathways of a finite length at a single temperature at a time. In contrast, ReactCA can explore pathways of unlimited length (and allow them to interact) over a range of temperatures in a single simulation. These improvements both simplify the analytic process and broaden the range of reaction behavior that can be predicted compared to the reaction network.

In addition to the reaction intermediates identified by Todd et al., the ReactCA simulation predicts a number of other unobserved intermediates, most of which appear in the simulation at only trace levels (less than 1-2\% by mass, as shown in Figure \ref{fig:ymno3_overall_rxn}b). These unobserved intermediates reveal the alternate pathways which are present in the trajectory as a result of the automaton sampling many available reactions during its evolution. While we demonstrated earlier that investigating even low-prevalence intermediates, such as \ce{LiMnO2}, can yield insights into overall reaction pathways, the prevalence of a given intermediate in a ReactCA simulation generally reflects the relative amount predicted to appear during the synthesis reaction. In other words, while we do show that \ce{LiMnO2} plays a role in the reaction that aligns well with the experimental hypothesis, its low prevalence suggests that it may not accumulate in significant quantities during the reaction. Similarly, because these other low-prevalence phases appear in only trace quantities, the ReactCA simulation should be interpreted as assigning low likelihood to the appearance of those phases in experimental characterization.

Besides these low-prevalence unobserved impurities, two others (\ce{YMn2O5} and \ce{Mn8Cl3O10}) achieve significant amounts comparable to the predictions for the observed intermediates (\textgreater10\% by mass). The first of these phases, \ce{YMn2O5}, does not appear in the experiment originally presented by Todd et al., but is recognized in a later work by the same authors as a common impurity in the synthesis of \ce{YMnO3} by this metathesis method \cite{todd_selectivity_2021}. The second of these phases, \ce{Mn8Cl3O10} was recently synthesized by the solid-state method from precursors \ce{MnCl2} and \ce{MnO2} \cite{dosaev_synthesis_2022} at 600C, hence its appearance is not implausible in this reaction. However, by examining the reactions which consume \ce{Mn8Cl3O10} during the ReactCA simulation, we find that its primary role is as an intermediate between reactants \ce{YCl3} and \ce{Mn2O3} (which react to form it), and experimentally verified downstream intermediates (most significantly, \ce{LiMnO2}, \ce{Mn3O4}, and \ce{LiCl}). As a result, we hypothesize that \ce{Mn8Cl3O10} functions similarly to \ce{ZrN} in our discussion of the \ce{CaZrN2} synthesis above. That is, \ce{Mn8Cl3O10} may be the best representation the simulation can provide of a process which is actually facilitated by highly defective or amorphous intermediates with a similar Mn-Cl-O composition that are not present among the ordered, crystalline, stoichiometrically exact phases on the Materials Project. This uncertainty further emphasizes the necessity of future work to improve the ability of the automaton to navigate more sophisticated intermediate landscapes.

\section{Conclusions}

We present ReactCA, a new simulation framework based on the cellular automaton formalism for predicting the evolution of crystalline phases during the course of a solid-state reaction. This simulation utilizes thermodynamic data from the Materials Project and machine learning estimated melting points in conjunction with a cost function to assign reaction rates as a function of temperature. The evolution of the reacting material is determined by a rule based on the pairwise interface reaction model that considers reactions between neighboring particles. The flexibility of the form of both the cost function and the evolution rule lend great extensibility, meaning that as new data become available from as of yet unrealized high-throughput methods or machine learning frameworks, both empirical rules and heuristics based on those data can be incorporated into ReactCA to improve its performance. We illustrate the current performance of the simulation framework using three case study systems, the first of which serves as a platform for viewing the relative selectivity of various reaction recipes and the importance of precursor choice with regard to product purity, and the second and third of which demonstrate power of ReactCA in predicting likely reaction pathways, the temperature dependence of those pathways, and the order and amounts of intermediate phases which appear during the course of complex ternary metathesis reactions.

While we believe that ReactCA is of immediate utility in both ``testing'' reaction recipes before utilizing physical or monetary resources to experimentally execute them, we also foresee the simulation being used to determine recipe parameters using an optimization framework or to facilitate the autonomous design of synthesis recipes by acting as a digital twin for experimental synthesis in an automated lab. Still, there are many physical phenomena that are not considered by ReactCA. In particular, future work that will most significantly improve this simulation is related to the intelligent incorporation of kinetic effects into the evolution rule and source data. The case studies presented in this work show that the current model likely overpredicts the number of ordered intermediates or impurity phases as compared to real synthesis experiments. Furthermore, including complex, non-crystalline intermediates such as off-stoichiometric and amorphous phases as well as new phase transition mechanisms (e.g. sublimation or peritectic decomposition) provide future challenges. We foresee ReactCA growing alongside increasingly rich data generation capabilities in computational materials science to capture improved features of solid-state synthesis as the field matures, yielding faster and more accurate methods for predicting solid-state synthesis behavior.

\section{Methods}

\subsection{Thermochemistry Data}

To prepare data for the simulation of the \ce{BaTiO3}, \ce{CaZrN2} and \ce{YMnO3} recipes, entries from the Materials Project from the Ba-Ti-O-C, Ba-Ti-O-Na-S, Ba-Ti-O-Na-Cl, Ca-Zr-N-Cl, and Y-Mn-Cl-Li-C-O chemical systems were collected. Additionally, a density functional theory structure relaxation calculation was performed at the GGA level of theory using parameters from the \verb|MPRelaxSet| in \verb|atomate2|\cite{ganose_atomate2_2024} to obtain a formation enthalpy for \ce{Y3O4Cl}, a phase known to appear during the synthesis of \ce{YMnO3}, but not present within the Materials Project. For each entry in these chemical systems, the machine learning descriptor by Bartel et al. was used to estimate the vibrational entropy contribution to the Gibbs energy of formation at 300K \cite{bartel_physical_2018}. Phases with formation energies greater than 30 meV/atom above the hull were removed at this temperature. This value for a metastability cutoff filter was chosen based in part on the work of Sun et al., which present statistics on the metastability of compounds in the Materials Project and connects them to synthesizability \cite{sun_thermodynamic_2016}. We also removed compounds which were marked as ``theoretical'' on the Materials Project unless they appeared explicitly in the experimental results. In addition, several phases which have previously only been successfully produced using synthesis methods other than the solid-state method considered here were excluded. These are \ce{ZrCl} (synthesized using high pressure methods \cite{adolphson_crystal_1976}), \ce{ZrCl2} (synthesized via hydrogen disproportionation reactions \cite{imoto_synthesis_1981}), \ce{CaN2} (synthesized using high pressure methods \cite{schneider_synthesis_2012}), \ce{CaN6} (synthesized in solution \cite{krischner_kristallstrukturbestimmung_1982}), \ce{Ca2N} (synthesized via metallothermic reduction \cite{reckeweg_alkaline_2002}), and \ce{Zr3N4} (synthesized via high pressure methods \cite{taniguchi_synthesis_2019, zerr_synthesis_2003} or ammonolysis \cite{lerch_synthesis_1996}). 

Data from the Materials Project was collected using the \verb|mp-api| package and the possible reactions in each of these systems were enumerated using the \verb|reaction-network| Python package \cite{mcdermott_graph-based_2021}. The CA model was implemented using the \verb|pylattica| Python package \cite{gallant_pylattica_2024}.

\section{Data Availability}

Enumerated reactions and their scores, recipe documents, and serialized simulation results are provided in the Supplementary Information.

\section{Code Availability}

The implementation for this automaton is available at \url{https://github.com/mcgalcode/rxn-ca}. As implemented in the automaton repository, reaction enumeration was performed using the \verb|reaction-network| package available at \url{https://github.com/materialsproject/reaction-network}.

\section{Acknowledgements}

This work was primarily financed by the U.S. Department of Energy, Office of Science, Office of Basic Energy Sciences, Materials Sciences and Engineering Division under contract no. DE-AC02-05-CH11231 (D2S2 programme, KCD2S2) and the Laboratory Directed Research and Development Program of Lawrence Berkeley National Laboratory. This research used resources of the National Energy Research Scientific Computing Center (NERSC), a U.S. Department of Energy Office of Science User Facility operated under Contract No. DE-AC02-05CH11231.

\subsection{Author Contributions}

\textbf{Max C. Gallant}: conceptualization (lead); software (lead); writing - original draft (lead). \textbf{Matthew J. McDermott}: conceptualization (supporting); writing - review and editing (equal). \textbf{Bryant Li}: conceptualization (supporting); writing - review and editing (supporting). \textbf{Kristin A. Persson}: supervision; funding acquisition; writing - review and editing (equal).

\bibliographystyle{nature_adk}
\bibliography{rxn_ca}
\input{si}

\end{document}

%% file: si.tex
\clearpage

\setcounter{page}{1}
\setcounter{section}{0}
\setcounter{table}{0}
\setcounter{figure}{0}
\setcounter{equation}{1}

\renewcommand{\thepage}{S\arabic{page}}
\renewcommand{\thesection}{S\arabic{section}}
\renewcommand{\theequation}{S\arabic{equation}}
\renewcommand{\thetable}{S\arabic{table}}
\renewcommand{\thefigure}{S\arabic{figure}}

\clearpage

\begin{center}
\subsection*{Supplemental Material:\\
ReactCA: A Cellular Automaton for Predicting Phase Evolution in Solid-State Reactions}    
\end{center}

\pagebreak

\section{Swapping probability as a function of temperature}
\begin{figure}[ht!]
\centering
\includegraphics[width=0.7\textwidth,keepaspectratio]{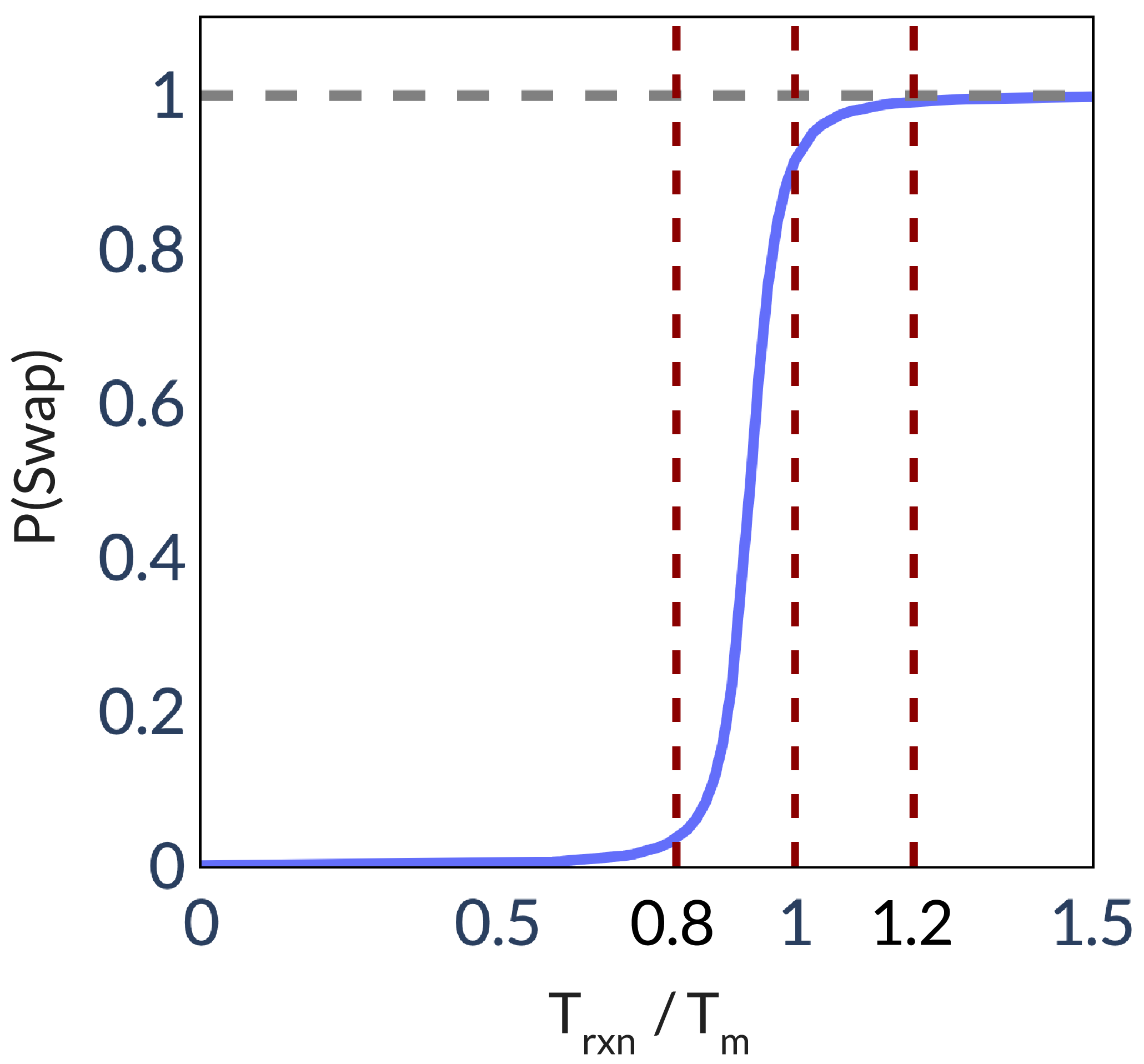}
\caption{\label{supp_fig:swap_probability}\textbf{Swapping probability during the Melt Swap  action} The probability of neighboring cells swapping during an application of the evolution rule is a function of the ratio of the melting point of the phase occupying the evolving cell and the current reaction temperature. The onset of swapping behavior occurs at 80\% of the melting temperature of the phase in the evolving cell, and reaches 100\% by the time the reaction temperature exceeds this melting point by 20\%.}
\end{figure}

\pagebreak

\section{Description of the Reaction Progression action}\label{sec:reaction_progression}

The Reaction Progression evolution rule action is designed such that mass is conserved when it is applied repeatedly. This conservation is achieved by the sufficient sampling of probability distributions that are constructed based on the stoichiometry of the reaction. An example presented here illustrates this process.

Consider two neighboring cells, denoted $C_1$ and $C_2$, the contents of which  are undergoing a reaction. $C_1$ is occupied by phase A and $C_2$ is occupied by phase B. Assume the following reaction (whose coefficients are expressed in units of volume) has been selected to proceed between these cells:
\begin{equation}
    \label{supp_equation:reaction}
    A + 2B \rightarrow 3C + 4D
\end{equation}

After the reaction is selected to proceed, the new state of each of these two cells is determined independently.

Whether or not the contents of each cell are consumed is determined by treating the coefficients of the reactants in (\ref{supp_equation:reaction}) as frequencies. In other words, when this reaction proceeds, it consumes A one third of the time and B two thirds of the time. To capture this, we begin by constructing a distribution over the reactants using these coefficients as weights. Formally, for reaction $T$ with reactants $R$ and products $P$ and stoichiometric coefficients $r$ and $p$, the probability that reactant $r_i$ is consumed is given by the following: 
\begin{equation}
    P(R_{i}\text{ is consumed}) = \frac{r_i}{\sum_{j \in r} r_j}
\end{equation}
\begin{equation}
    P(R_{i}\text{ is not consumed}) = \frac{\sum_{j \in r, j \neq i} r_j}{\sum_{j \in r} r_j}
\end{equation}
Applying this to the reaction in (\ref{supp_equation:reaction}), the following probabilities are generated for $C_1$:
\begin{align*}
    P(\text{A is consumed}) = \frac{1}{1 + 2} = \frac{1}{3} &&
    P(\text{A is not consumed}) = \frac{2}{1 + 2} = \frac{2}{3}
\end{align*}
And the following are generated for $C_2$:
\begin{align*}
    P(\text{B is consumed}) = \frac{2}{1 + 2} = \frac{2}{3} &&
    P(\text{B is not consumed}) = \frac{1}{1 + 2} = \frac{1}{3}
\end{align*}

For each cell, a random draw from the corresponding distribution is performed to determine whether or not the contents are consumed. As illustrated above, there is a 1/3 chance that $C_1$ (which contains phase A) is consumed and a 2/3 chance that it is not. Similarly, there is a 2/3 chance that $C_2$ (which contains phase B) is consumed, and a 1/3 chance that it is not. These draws are performed independently.

Next, for each of the cells that is consumed (it might have been one of them, both of them, or neither of them), a product phase is selected. To make this selection we again treat the reaction coefficients as frequencies. Again considering reaction $T$ involving reactants $R$ (with coefficients $r$) and products $P$ (with coefficients $p$), the probability of producing of producing phase $P_i$ is given as follows: 
\begin{equation}
    P(P_{i}\text{ is produced}) = \frac{p_i}{\sum_{j \in p} p_j}
\end{equation}
For the reaction in (\ref{supp_equation:reaction}), these probabilities work out to be:
\begin{align*}
    P(\text{C is produced}) = \frac{3}{3 + 4} = \frac{3}{7} &&
    P(\text{D is produced}) = \frac{4}{3 + 4} = \frac{4}{7}
\end{align*}

Using these likelihoods, at each consumed cell, a random draw is performed, and the content of the cell is replaced by the phase resulting from the draw.

It is important to note that reactions do not conserve volume. In this example, the reaction consumes 3 units of volume and produces 7 units of volume. To capture this, the volume of each consumed cell is scaled by this ratio:

\begin{equation}
    \label{supp_equation:volume_scaling}
    V_{new} = \frac{V_{products}}{V_{reactants}} * V_{old}
\end{equation}

Finally, the simulation state is updated with the new volume and product phase for each cell which was consumed by the reaction (if any), and the next application of the evolution rule begins.

\newpage

\newpage
\section{Outcome convergence}
\begin{figure}[ht!]
\centering
\includegraphics[width=0.7\textwidth,keepaspectratio]{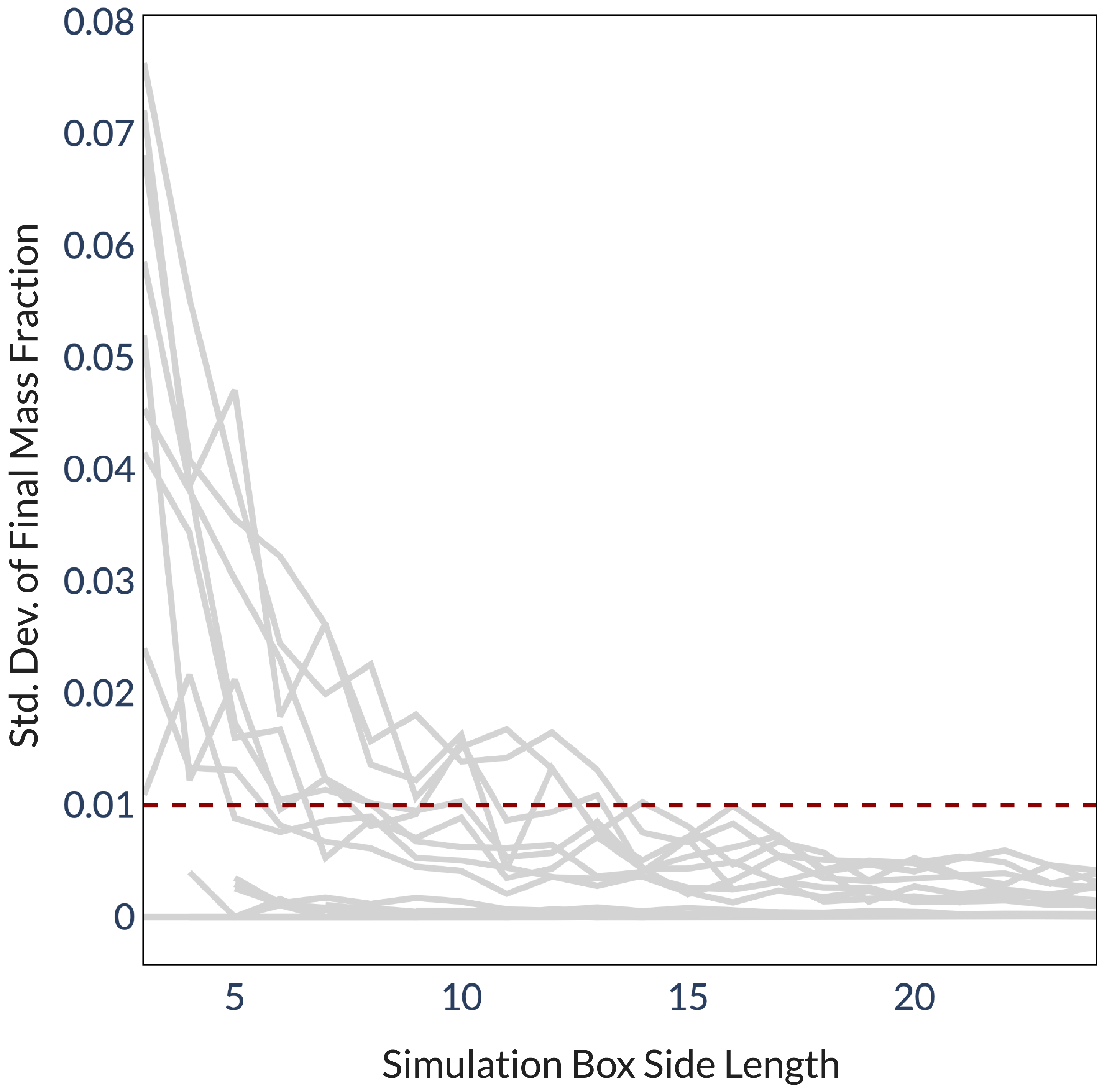}
\caption{\label{supp_fig:convergence_final_mass}\textbf{Convergence of the standard deviation of final mass fraction for each phase} For each in a range of simulation sizes (side lengths between 3 and 24), 10 simulations were run. For each phase that appeared, the mass fraction attained at the end of each simulation was calculated. For each simulation size, the standard deviation of these final mass fraction values was calculated on a per phase basis. The plot above shows the convergence behavior of these standard deviations on a per phase basis (one trace per phase). Notably, for simulation boxes with side length greater than 15 cells, the standard deviation of the final mass fraction of each phase over repeated simulations is less than 1\%.}
\end{figure}

\newpage

\begin{figure}[ht!]
\centering
\includegraphics[width=0.7\textwidth,keepaspectratio]{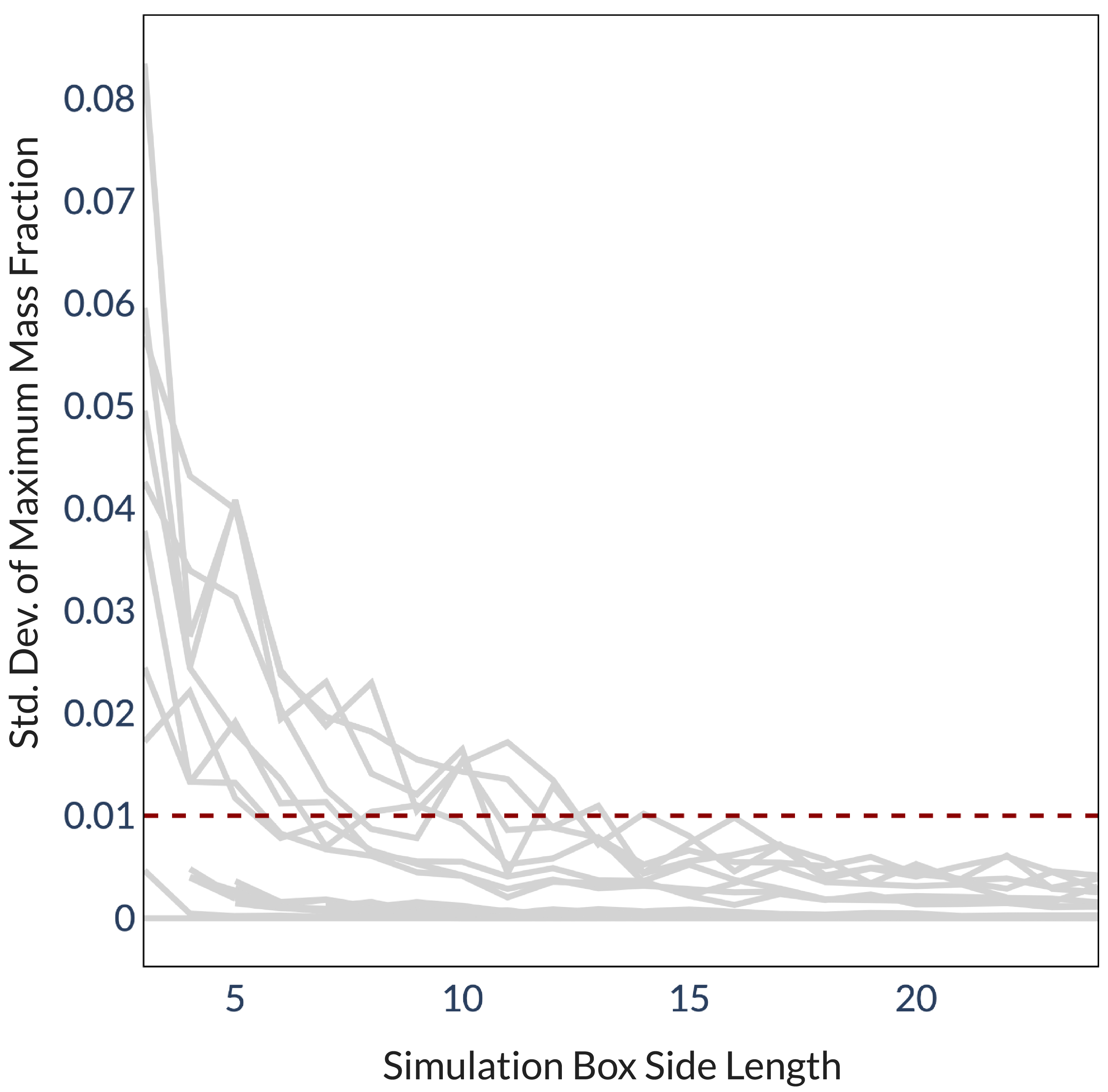}
\caption{\label{supp_fig:convergence_max_mass}\textbf{Convergence of the standard deviation of maximum mass fraction achieved for each phase} For each in a range of simulation sizes (side lengths between 3 and 24), 10 simulations were run. For each phase that appeared, the maximum mass fraction attained at any point during each simulation was calculated. For each simulation size, the standard deviation of these maximum mass fraction values was calculated on a per phase basis. The plot above shows the convergence behavior of these standard deviations on a per phase basis (one trace per phase). Notably, for simulation boxes with side length greater than 15 cells, the standard deviation of the maximum mass fraction of each phase over repeated simulations is less than 1\%.}

\end{figure}

\newpage

\begin{figure}[ht!]
\centering
\includegraphics[width=\textwidth,keepaspectratio]{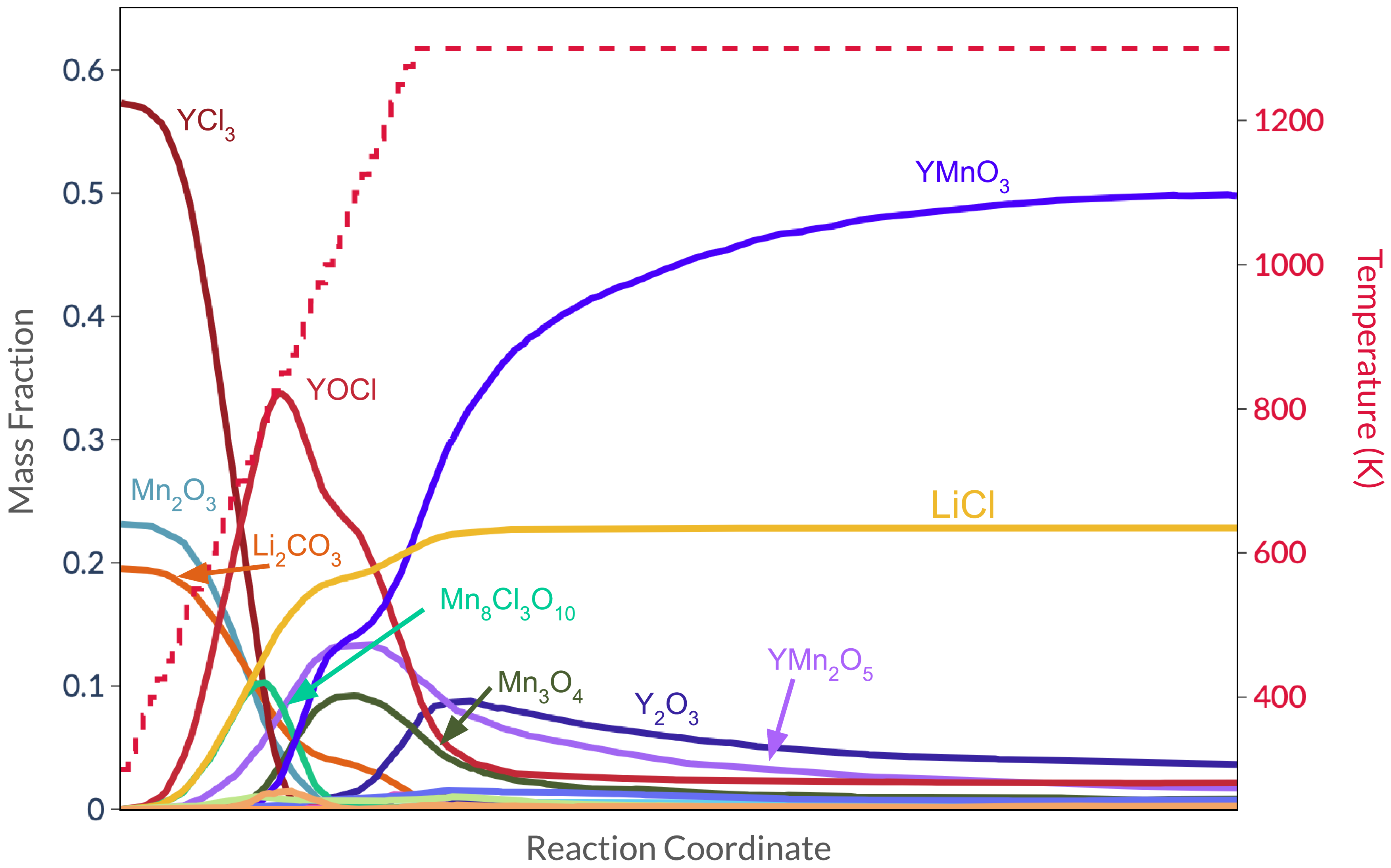}
\caption{\label{supp_fig:long_ymno3_traj}\textbf{Extended trajectory for \ce{YMnO3} reaction} A longer trajectory for the \ce{YMnO3} reaction showing the plateauing of each of the eventual products.}

\end{figure}